\documentclass[prd,twocolumn,superscriptaddress,showpacs,preprintnumbers,nofootinbib,APS]{revtex4}
\usepackage{color}
\usepackage{psfrag}
\usepackage{dcolumn}
\usepackage{bm}
\usepackage[latin1]{inputenc}
\usepackage[spanish,english]{babel}
\usepackage{amsfonts}
\usepackage{amssymb,epsf}
\usepackage{graphicx}
\usepackage{epstopdf}

\begin{document}

\title{\textbf{\ van der Waals like behaviour of topological AdS black holes
\\ in massive gravity }}
\author{S. H. Hendi}
\affiliation{{\footnotesize Physics Department and Biruni
Observatory, College of Sciences, Shiraz University, Shiraz 71454,
Iran}} \affiliation{{\footnotesize Research Institute for
Astronomy and Astrophysics of Maragha (RIAAM), Maragha, Iran}}
\author{R. B. Mann}
\affiliation{{\footnotesize Department of Physics and Astronomy,
University of Waterloo, 200 University Avenue West, Waterloo,
Ontario, Canada N2L 3G1}}
\author{S. Panahiyan}
\affiliation{{\footnotesize Physics Department and Biruni
Observatory, College of Sciences, Shiraz University, Shiraz 71454,
Iran}} \affiliation{{\footnotesize Physics Department, Shahid
Beheshti University, Tehran 19839, Iran }}
\author{B. Eslam Panah}
\affiliation{{\footnotesize Physics Department and Biruni
Observatory, College of Sciences, Shiraz University, Shiraz 71454,
Iran}}

\begin{abstract}
Motivated by recent developments in black hole thermodynamics, we
investigate van der Waals phase transitions of charged black holes
in massive gravity. We find that massive gravity theories can
exhibit strikingly different thermodynamic behaviour compared to
that of Einstein gravity, and that the mass of the graviton can
generate a range of new phase transitions for topological black
holes that are otherwise forbidden.
\end{abstract}

 \maketitle



\noindent \textit{Introduction-}  Understanding the quantum
behaviour of gravity could be related to the possible mass of the
graviton. The consistency of including this in the context of
extending general relativity has been a long-standing basic
physical question of classical field theory. Although the
primitive linear theory of massive gravity \cite{Fierz} contains
Boulware-Deser ghost modes \cite{Boulware}, a nonlinear
generalization that is ghost free to all orders has recently been
constructed \cite{dRGT} by de Rham, Gabadadze and Tolley (dRGT
massive
theory). dRGT is stable, enjoys the absence of Boulware-Deser ghost \cite%
{HassanR}, and has yielded a number of intriguing results in terms
of its cosmological behaviour and black hole solutions
\cite{Fasiello}. The mass terms are produced by consideration of a
reference metric, which plays a crucial role in the construction
of dRGT \cite{de RhamLive}. Motivated by applications of
gauge/gravity duality, Vegh proposed a new reference metric in
which the graviton  behaves  like a lattice excitation and
exhibits a Drude peak \cite{Vegh}. This theory is also ghost-free
and stable \cite{Zhang}, and $d-$dimensional ($d\geq 3$) black
hole solutions  in the presence of linear and nonlinear
electrodynamics with van der Waals like behavior have been
obtained \cite{CaiH}. Higher curvature generalizations have also
been constructed \cite{HendiPE}.  Although some classes of
nonlinear massive gravity theories are Lorentz-violating and bear
a close relation to Horava-Lifshitz gravity \cite{Hansen}, it was
shown that there are Lorentz invariant versions of nonlinear
massive gravity  as well \cite{dRGT}.  Massive
gravity is also motivated by observation. Obtaining an empirical
upper limit on the mass of the graviton remains an outstanding
challenge (for more details see \cite{gravitonMASS}) , one that
should soon be attainable once recent LIGO results \cite{LIGO} are
improved and expanded. In this regard, one may use the results of
Refs. \cite{Finn} and \cite{Jimenez} to obtain a bound on the
energy flux emitted from a binary pulsar and on the propagation
speed of the graviton.

Here, we consider a class of dRGT theories, which we regard as the
minimal modification of general relativity that yields a massive
graviton \cite{dRGT}. We demonstrate that black holes of
non-spherical topology can exhibit van der Waals phase transitions
in dRGT like gravity. Such transitions are forbidden in standard
Einstein gravity  and also higher curvature theories such as
Lovelock gravity.

The study of black hole thermodynamics began with the pioneering
work of the Hawking and Page \cite{HawkingPage} that indicated
anti-de Sitter (AdS) black holes can undergo phase transitions.
Asymptotically AdS black holes have been of special interest since
they admit a gauge/gravity duality description, and their
thermodynamics plays a crucial role in constructing a consistent
theory of quantum gravity \cite{ADSCFT}. Indeed, this duality can
be applied to a qualitative study of the behaviour of various
condensed matter phenomena \cite{Duality}. Substantial progress
was recently made when van der Waals behaviour of asymptotically
AdS charged black holes was observed \cite{VdW}. Based on the
canonical ensemble, a small-large AdS black hole phase transition
analogous to the liquid-gas phase transition in a thermodynamic
system was discovered. A number of significant results were
subsequently obtained, including the existence of triple points \cite{triple}%
, reentrant phase transitions \cite{reent}, and analogous
Carnot-cycle heat engines \cite{heat}. These properties
established a connection between black hole thermodynamics and
everyday ``chemical" thermodynamics, known as ``black hole
chemistry". It likewise trigged investigations into the
implications for gauge/gravity duality, such as holographic
superconductors \cite{super}, the Kerr/CFT correspondence
\cite{ker}, and holographic entanglement entropy \cite{entalg}
(see \cite{Mann} for a review).

However, van der Waals behaviour (and its applications) in Einstein gravity
are seen only for AdS black holes with spherical horizon topology \cite{VdW}%
; no such behaviour takes place for AdS black holes with flat or
hyperbolic horizons (no real critical point). Such reports
motivate one to look for an extension of Einstein gravity to a
modified version, in which van der Waals behaviour is seen not
only for spherical AdS black holes, but for AdS black holes with
different horizon curvature. In what follows, we demonstrate that
black holes in massive gravity exhibit van der Waals behaviour
independent of the choice of horizon curvature. This relaxes the
constraint on the topological structure of the solutions, allowing
the possibility of conducting studies in the context of black hole
chemistry, the AdS/CFT correspondence, and the non-relativistic
AdS/CMT correspondence, regardless of horizon geometry.


\noindent \textit{Basic Setup-} The $d$-dimensional action of Einstein-$%
\Lambda $-massive gravity with a $U(1)$ gauge field is
\begin{equation}
\mathcal{I}=\frac{-1}{16\pi }\int d^{d}x\sqrt{-g}\left( \mathcal{R}-2\Lambda
-\mathcal{F}+m^{2}\sum_{i}^{4}c_{i}\mathcal{U}_{i}(g,f)\right) ,
\label{Action}
\end{equation}%
in which $\mathcal{R}$ is the scalar curvature of the metric $g_{\mu\nu}$, $%
\Lambda $ is the negative cosmological constant and $\mathcal{F}=F_{\mu \nu
}F^{\mu \nu }$ is the Maxwell invariant, where $F_{\mu \nu }=\partial _{\mu
}A_{\nu }-\partial _{\nu }A_{\mu }$ is the electromagnetic tensor with gauge
potential $A_{\mu }$, $m$ is massive parameter, and $f_{\mu\nu}$ is a fixed
symmetric tensor. The $c_{i}$'s are constants and the $\mathcal{U}_{i}$'s
are symmetric polynomials of the eigenvalues of the $d\times d$ matrix $%
\mathcal{K}_{\nu }^{\mu }=\sqrt{g^{\mu \alpha }f_{\alpha \nu }}$, where
\begin{eqnarray*}
\mathcal{U}_{1}&=&\left[ \mathcal{K}\right], \;\;\;\;\;\;\;\;\;\;\; \mathcal{%
U}_{2}=\left[ \mathcal{K}\right] ^{2}-\left[ \mathcal{K}^{2}\right] , \\
\mathcal{U}_{3}&=&\left[ \mathcal{K}\right] ^{3}-3\left[ \mathcal{K}\right] %
\left[ \mathcal{K}^{2}\right] +2\left[ \mathcal{K}^{3}\right], \\
\mathcal{U}_{4}&=&\left[ \mathcal{K}\right] ^{4}-6\left[ \mathcal{K}^{2}%
\right] \left[ \mathcal{K}\right] ^{2}+8\left[ \mathcal{K}^{3}\right] \left[
\mathcal{K}\right] +3\left[ \mathcal{K}^{2}\right] ^{2}-6\left[ \mathcal{K}%
^{4}\right].
\end{eqnarray*}

Variation of the action (\ref{Action}) with respect to the metric tensor, $%
g_{\mu \nu }$, and the Faraday tensor, $F_{\mu \nu }$, leads to
\begin{eqnarray}
G_{\mu \nu }+\Lambda g_{\mu \nu }+m^{2}\chi _{\mu \nu } &=&-2\left( F_{\mu
\rho }F_{\nu }^{\rho }-\frac{1}{4}g_{\mu \nu }\mathcal{F}\right) ,
\label{Field equation} \\
\partial _{\mu }\left( \sqrt{-g}F^{\mu \nu }\right) &=&0,
\label{Maxwell equation}
\end{eqnarray}%
in which $G_{\mu \nu }$ is the Einstein tensor and $\chi _{\mu \nu }$ is
\begin{eqnarray}
&&\chi _{\mu \nu } = \frac{c_{1}}{2}\left( \mathcal{K}_{\mu \nu }-\mathcal{U}%
_{1}g_{\mu \nu }\right)- \frac{c_{2}}{2}\left( \mathcal{U}_{2}g_{\mu \nu }-2%
\mathcal{U}_{1}\mathcal{K}_{\mu \nu }+2\mathcal{K}_{\mu \nu }^{2}\right)
\nonumber \\
&&-\frac{c_{3}}{2}(\mathcal{U}_{3}g_{\mu \nu }-3\mathcal{U}_{2}\mathcal{K}%
_{\mu \nu }+6\mathcal{U}_{1}\mathcal{K}_{\mu \nu }^{2}-6\mathcal{K}_{\mu \nu
}^{3})-\frac{c_{4}}{2} \times  \nonumber \\
&& \left(\mathcal{U}_{4}g_{\mu \nu }-4\mathcal{U}_{3}\mathcal{K}_{\mu \nu }+
12\mathcal{U}_{2}\mathcal{K}_{\mu \nu }^{2}-24\mathcal{U}_{1} \mathcal{K}%
_{\mu \nu }^{3}+24\mathcal{K}_{\mu \nu }^{4} \right).
\end{eqnarray}

In order to obtain AdS topological static charged black holes in massive
gravity, we consider the metric of $d=(n+2)$-dimensional spacetime in the
following form
\begin{equation}
ds^{2}=-\psi \left( r\right) dt^{2}+\frac{dr^{2}}{\psi \left( r\right) }%
+r^{2}h_{ij}dx_{i}dx_{j},  \label{Metric}
\end{equation}%
where $i,j=1,2,3,...,n$ and $h_{ij}dx_{i}dx_{j}$ is a spatial metric of
constant curvature $d_{2}d_{3}k$ and volume $\mathcal{V}_{d_{2}}$, where $%
d_{i}=d-i$. The reference metric $f_{\mu \nu }$ is related to the
spatial components $h_{ij}$ of the spacetime line element.
Accordingly, we employ the ansatz $f_{\mu \nu
}=\text{diag}(0,0,c^{2}h_{ij})$ yielding
\begin{equation}
\mathcal{U}_{j}=\frac{c^{j}}{r^{j}}\Pi _{k=2}^{j+1}d_{k},
\end{equation}%
where $c$ is a positive constant \cite{Cai2015}. This choice of reference
metric preserves general covariance in the temporal and radial coordinates,
but not in the transverse spatial coordinates \cite{Vegh} and so graviton
mass terms will have a Lorentz-breaking property.

Setting $d=4$, the gauge potential ansatz $A_{\mu }=h(r)\delta
_{\mu }^{0}$ yields from (\ref{Maxwell equation})
$F_{tr}=\frac{q}{r^{2}}$ as the only nonzero component of
electromagnetic field tensor, in which $q$ is an integration
constant and is related to the electrical charge.

The field equations then yield
\begin{equation}
\psi \left( r\right) =k-\frac{m_{0}}{r}-\frac{\Lambda r^{2}}{3}+\frac{q^{2}}{%
r^{2}}+m^{2}\mathcal{A},  \label{f(r)}
\end{equation}%
where $\mathcal{A}=\frac{cc_{1}}{2}r+c^{2}c_{2}+\frac{c^{3}c_{3}}{r}$. The
quantity $m_{0}$ is an integration constant that is related to the total
mass of this black hole. We note that for zero graviton mass ($m=0$), the
solution (\ref{f(r)}) reduces to the Reissner-Nordstr\"{o}m black hole in $4$%
-dimensions.
Calculations of the Ricci and Kretschmann scalars indicate a diverge at the
origin ($\lim_{r\rightarrow 0}R=\infty $ and $\lim_{r\rightarrow 0}R_{\alpha
\beta \gamma \delta }R^{\alpha \beta \gamma \delta }\longrightarrow \infty $%
); as ${r\longrightarrow \infty }$ we find $R_{\alpha \beta \gamma \delta
}\rightarrow \frac{\Lambda }{3}(g_{\alpha \gamma }g_{\beta \delta }-g_{\beta
\gamma }g_{\alpha \delta })$, confirming that the solution is asymptotically
AdS.

\begin{figure}[tbp]
$%
\begin{array}{cc}
\epsfxsize=4.35cm \epsffile{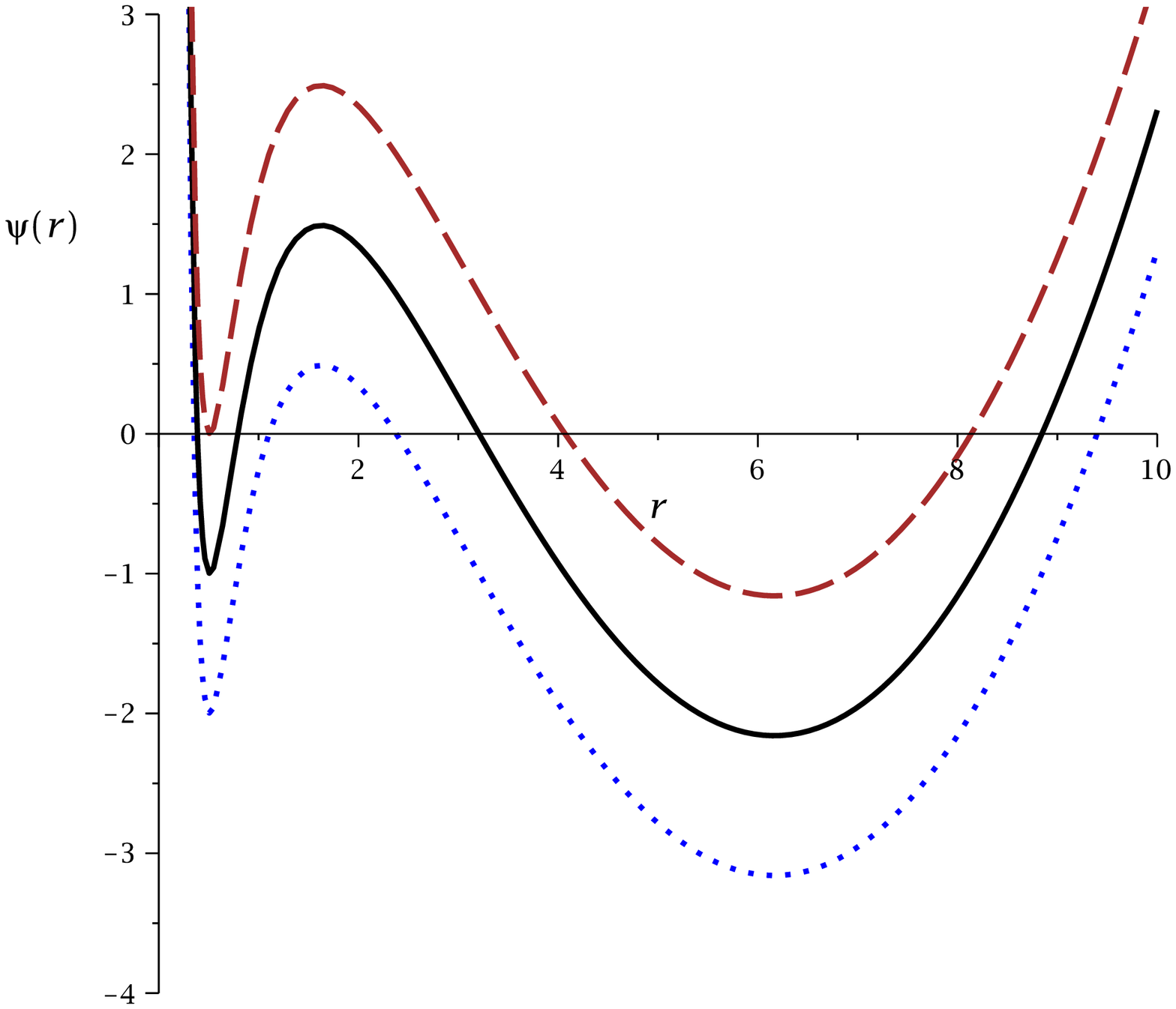} & \epsfxsize=4.35cm \epsffile{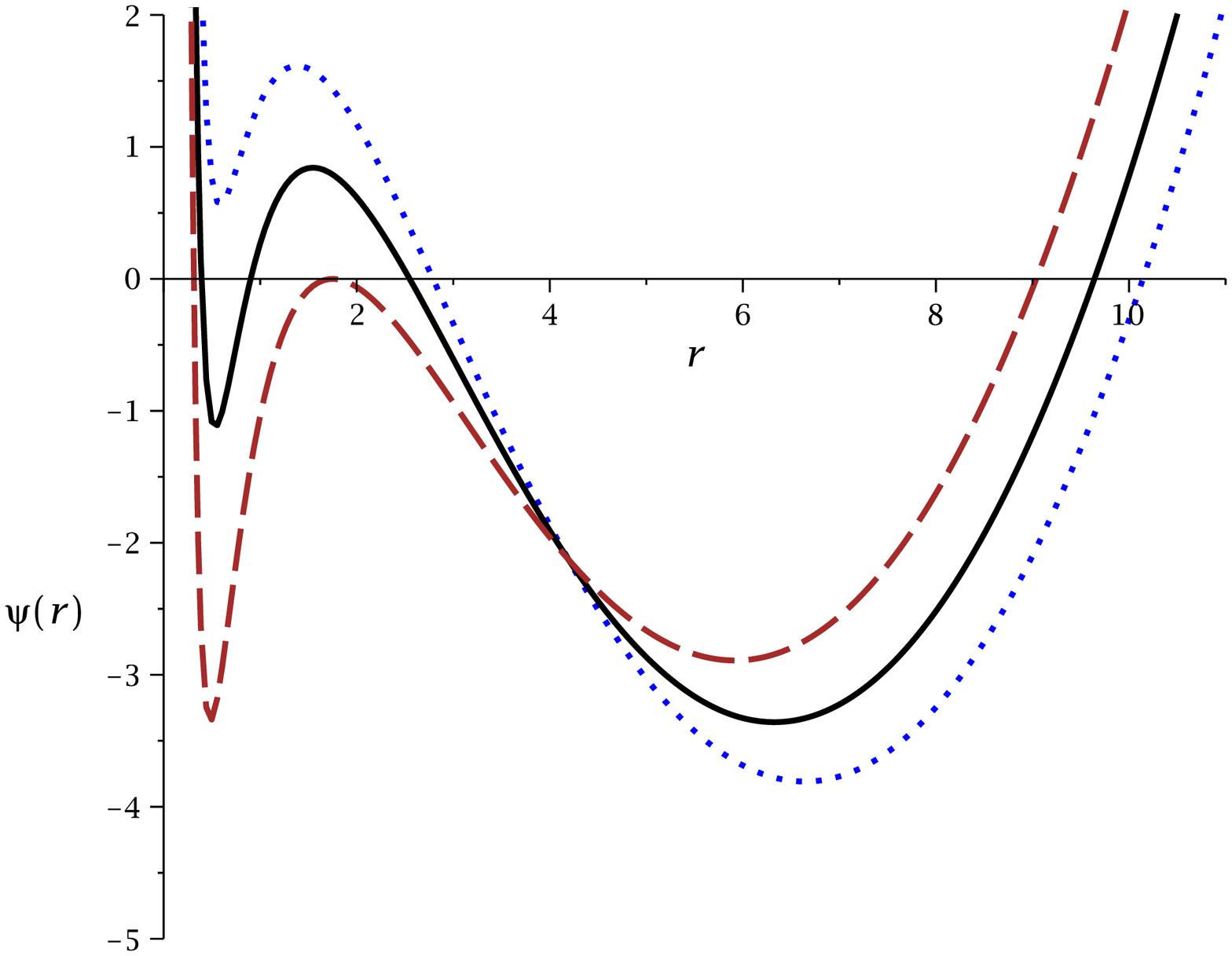}%
\end{array}
$%
\caption{$\protect\psi(r)$ versus $r$ for $\Lambda=-1$, $d=4$,
$q=1.7$, $c=1.00$, $c_{1}=-2.00$, $c_{2}=3.18$, $c_{3}=4.00$ and
$m_{0}=30$.  Left panel:  $m=2.1$, $k=-1$ (dotted line),
$k=0$ (continuous line) and $k=1$ (dashed line). Right
panel:  $k=-1$, $m=2.16$ (dotted line), $m=2.12$ (continuous
line) and $m=2.07$ (dashed line).} \label{Fig0}
\end{figure}

To study the effects of the massive terms on our solutions, we can
investigate the roots of the metric function ($\psi (r)=0$). In
massive gravity, it is possible for there to be as many as four
real roots in all horizon topologies: spherical ($k=1$), flat
($k=0$) and hyperbolic ($k=-1$), and we illustrate sample results
in table I  (see also Fig. \ref{Fig0}). The existence of more than
two roots for the metric function is due to the presence of the
massive terms.  Such multi-horizon solutions have been of interest
in understanding  anti-evaporation processes \cite{anti}. We
postpone a discussion of the causal and geodesic structures of
this class of solutions for future work, concentrating on their
thermodynamic behaviour.

\begin{table}[tbp]
\begin{center}
\begin{tabular}{ccc}
\hline\hline $k$ & $m$ & $roots$ \\ \hline \hline $1$ & $2.1600$ &
$3.8955,9.2446$ \\ \hline $0$\ \  & $2.1300$ & $3.1305,9.3969$ \\
\hline $-1$ & $2.1600$ & $2.6165,10.2691$ \\ \hline \hline $1$\ \
& $2.1059$ & $0.51778,3.8325,8.4374$ \\ \hline $0$ & $2.1291$ &
$0.5432,3.1271,9.3857$ \\ \hline $-1$ & $2.1530$ &
$0.5756,2.5755,10.1809$ \\ \hline \hline $1$ & $2.1000$ &\ \
$0.4381,0.6198,3.8237,8.3482$ \\ \hline $0$ & $2.1000$ &\ \
$0.3753,0.8565,2.9959,9.0022$ \\ \hline $-1$ & $2.1000$ &\ \
$0.3449, 1.2545, 2.1028, 9.5275$ \\ \hline \hline &  &
\end{tabular}%
\end{center}
\caption{roots of metric function ($\protect\psi (r)=0$) for
$\Lambda=-3$, $d=4$, $q=1.67$, $c=1.2$, $c_{1}=-2$, $c_{2}=2.2$,
$c_{3}=1.5$ and $m_{0}=34$.} \label{tab1}
\end{table}

\noindent \textit{Thermodynamics in the extended phase space-} In extended
phase space, the cosmological constant is regarded as a thermodynamic
variable corresponding to pressure, with $P=-\frac{\Lambda }{8\pi }$. This
postulate leads to an interpretation of the black hole mass as enthalpy \cite%
{Kastor:2009wy}. Using Gauss's law and counterterm methods, we compute the
various conserved and thermodynamic quantities of these solutions, obtaining
\begin{eqnarray}
T &=&\frac{k}{4\pi r_{+}}-\frac{r_{+}\Lambda }{4\pi
}-\frac{q^{2}}{4\pi r_{+}^{3}}  +\frac{m^{2}}{4\pi r_{+}}\left(
cc_{1}r_{+}+c_{2}c^{2}\right) ,
\label{TTO} \\
S &=&\frac{\mathcal{V}_{2}r_{+}^{2}}{4},\qquad
Q =\frac{\mathcal{V}_{2}q}{4\pi },\qquad M=\frac{\mathcal{V}_{2}m_{0}}{%
8\pi },  \label{QSTO}
\end{eqnarray}%
where $\mathcal{V}_{2}$ is the area of a unit volume of constant $(t,r)$
space ($4\pi $ for $k=0$). Also the electric potential is
\begin{equation}
\Phi =A_{\mu }\chi ^{\mu }\left\vert _{r\rightarrow \infty }\right. -A_{\mu
}\chi ^{\mu }\left\vert _{r\rightarrow r_{+}}\right. =\frac{q}{r_{+}}.
\label{TotalU}
\end{equation}

With these relations, we find that the solutions obey the first law of black
hole thermodynamics in an extended phase space (including massive variables)
\begin{equation}
dM=TdS+\Phi dQ+VdP+\mathcal{C}_{1}dc_{1}+\mathcal{C}_{2}dc_{2}+\mathcal{C}%
_{3}dc_{3},  \label{1stlaw}
\end{equation}%
where the conjugate quantities associated with the intensive parameters $S$,
$Q$, $P$, $c_{i}$'s are
\begin{eqnarray}
T &=&\left( \frac{\partial M}{\partial S}\right) _{Q,P,c_{i}},  \label{TU} \\
\Phi &=&\left( \frac{\partial M}{\partial Q}\right) _{S,P,c_{i}},  \\
V &=&\left( \frac{\partial M}{\partial P}\right) _{S,Q,c_{i}}=\frac{{%
\mathcal{V}_{2}}r_{+}^{3}}{3},  \label{Vol} \\
\mathcal{C}_{1} &=&\left( \frac{\partial M}{\partial c_{1}}\right)
_{S,Q,P,c_{2},c_{3}}=\frac{{\mathcal{V}_{2}}cm^{2}r_{+}^{2}}{16\pi },
\label{C1} \\
\mathcal{C}_{2} &=&\left( \frac{\partial M}{\partial c_{2}}\right)
_{S,Q,P,c_{1},c_{3}}=\frac{{\mathcal{V}_{2}}c^{2}m^{2}r_{+}}{8\pi },
\label{C2} \\
\mathcal{C}_{3} &=&\left( \frac{\partial M}{\partial c_{3}}\right)
_{S,Q,P,c_{1},c_{2}}=\frac{{\mathcal{V}_{2}}c^{3}m^{2}}{8\pi },  \label{C3}
\end{eqnarray}%
with $T$ and $\Phi $ given in Eqs. (\ref{TTO}) and (\ref{TotalU}). In
addition, the corresponding Smarr relation can be derived by a scaling
(dimensional) argument as
\begin{equation}
M=2TS-2PV+\Phi Q+\mathcal{C}_{1}c_{1}-\mathcal{C}_{3}c_{3},
\end{equation}%
where $c_{2}$ does not appear since it has scaling weight 0. Since the $c_{2}
$-term in the metric function is a constant term in four dimensions with no
thermodynamical contribution, we set $dc_{2}=0$.

We note that the thermodynamic volume (\ref{Vol}) does not depend on the
graviton mass. This in turn implies that the isoperimetric ratio
\begin{equation}
\mathcal{R}=\left( \frac{3{V}}{\mathcal{V}_{2}}\right) ^{\frac{1}{3}}\left(
\frac{\mathcal{V}_{2}}{{A}}\right) ^{\frac{1}{2}}=1,  \label{eq:ipe-ratio}
\end{equation}%
and so the reverse isoperimetric inequality ($\mathcal{R}\geq 1$) \cite%
{Cvetic:2010jb} can be satisfied.

\begin{figure}[tbp]
$%
\begin{array}{cc}
\epsfxsize=4cm \epsffile{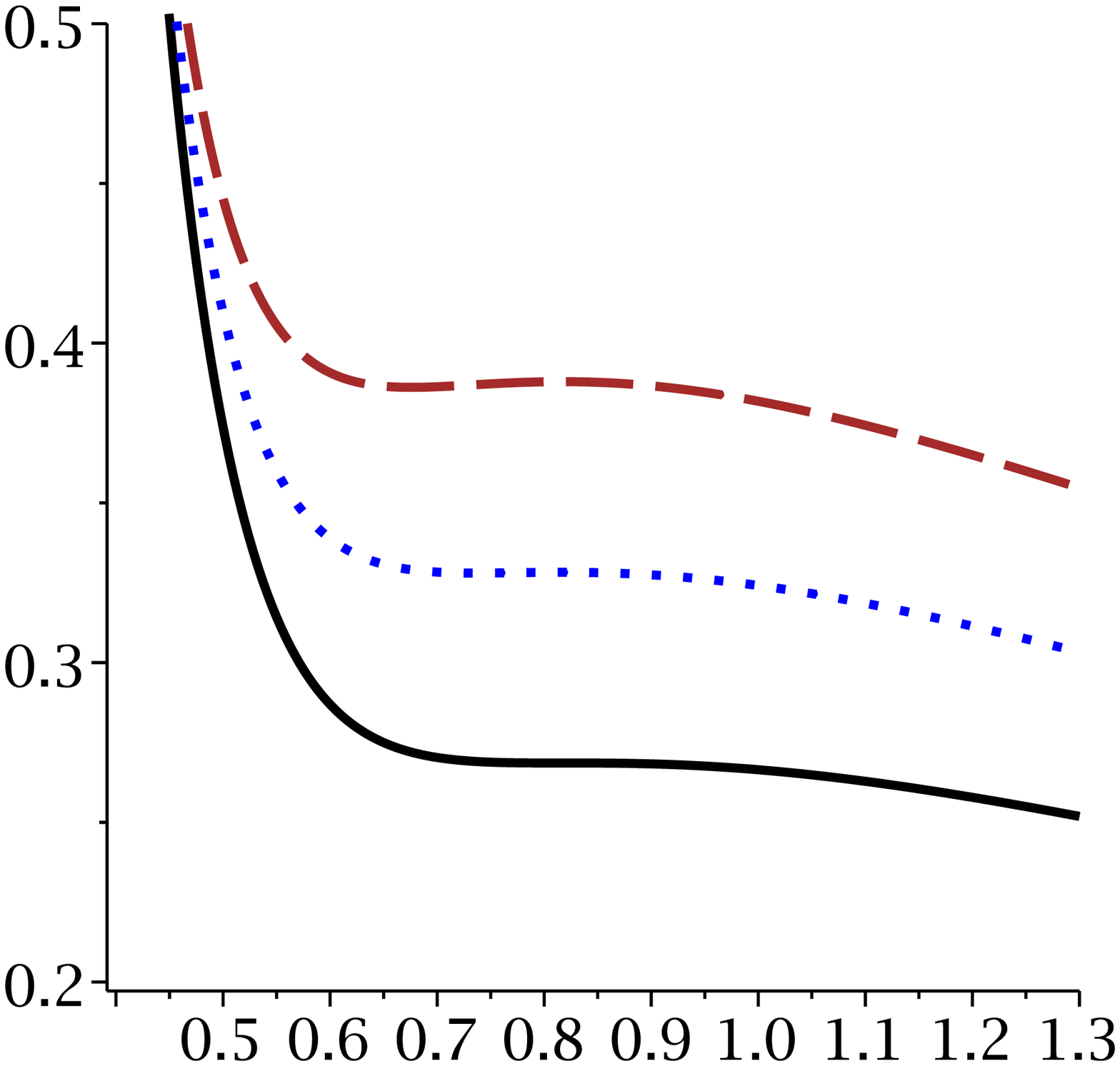} & \epsfxsize=4cm %
\epsffile{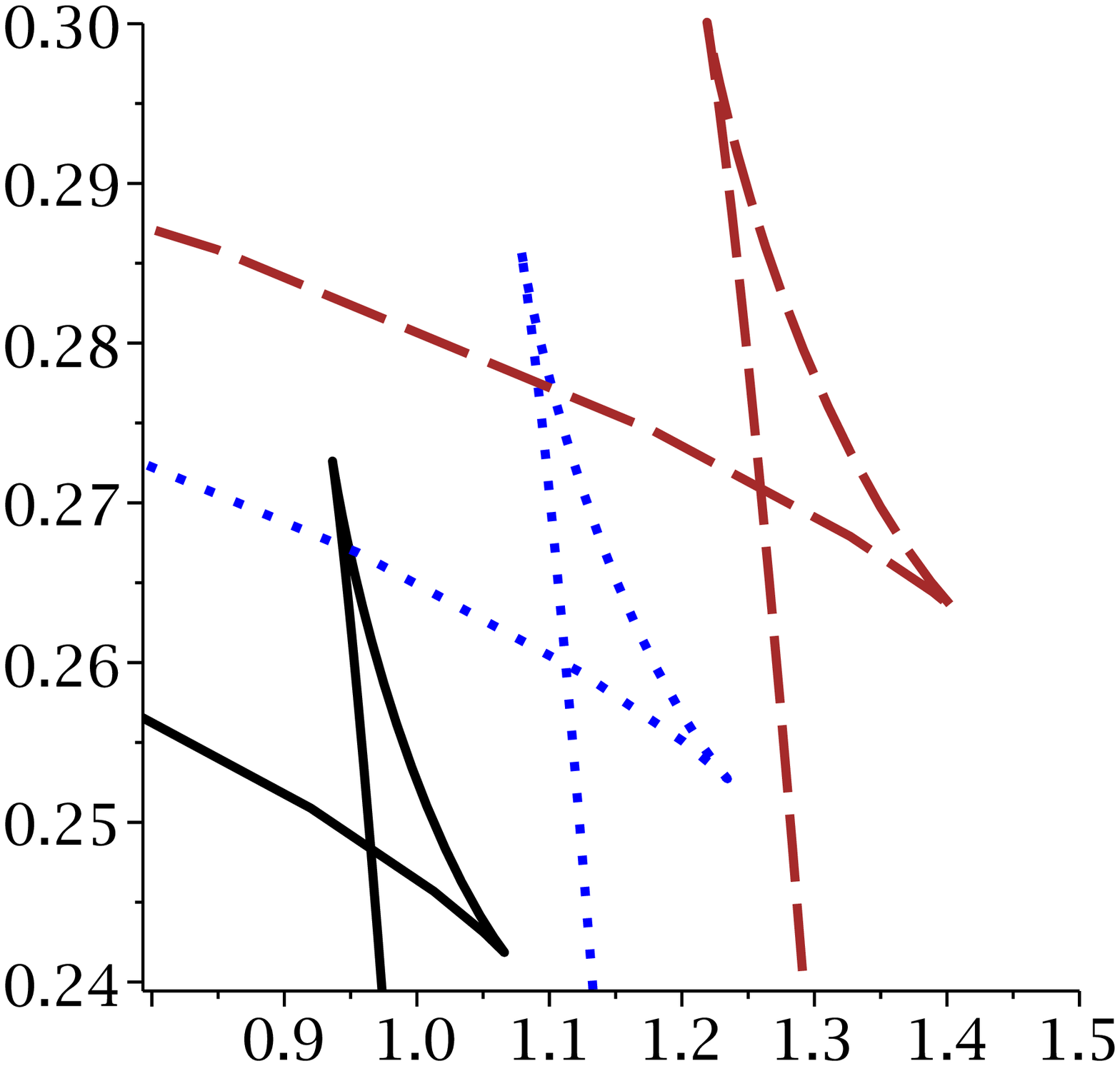}%
\end{array}
$%
\caption{$P-r_{+}$ (left panel) and $G-T$ diagrams for $d=4$,
$q=m=c=c_{1}=c_{3}=1$ and $c_{2}=10$. Left panel $T=T_{c}$, right
panel $P=0.5P_{c}$: $k=-1$ (continuous line), $k=0$ (dotted line)
and $k=1$ (dashed line).} \label{Fig1}
\end{figure}

To study critical phenomena and van der Waals behaviour, we compute the
equation of state and the Gibbs free energy
\begin{eqnarray}
P &=&\frac{4\pi T-m^{2}c_{1}c}{8\pi r_{+}}-\frac{k+m^{2}c_{2}c^{2}}{8\pi
r_{+}^{2}}+\frac{q^{2}}{8\pi r_{+}^{4}},  \label{EoS} \\
G &=&H-TS=M-TS=-\frac{Pr_{+}^{3}}{6}+  \nonumber \\
&&\frac{m^{2}r_{+}\left( c_{2}c^{2}r_{+}+2c_{3}c^{3}\right) +\left(
kr_{+}^{2}+3q^{2}\right) }{16\pi r_{+}},  \label{GG}
\end{eqnarray}%
using Eqs. (\ref{TTO}-\ref{QSTO}). Computing the inflection point $\left(
\frac{\partial P}{\partial r_{+}}\right) _{T}=\left( \frac{\partial ^{2}P}{%
\partial r_{+}^{2}}\right) _{T}=0$ of the equation of state, we find
\begin{equation}
 kr_{+c}^{2}-6q^{2}+m^{2}c_{2}c^{2}r_{+c}^{2}=0. \label{rc}
\end{equation}%
where $r_{+c}$ yields the critical volume $V_{c}$ via Eq. (\ref{Vol}). This
leads to the following respective critical horizon radius, temperature and
pressure
\begin{eqnarray}
r_{+c} &=&\frac{\sqrt{6}|q|}{\sqrt{k+m^{2}c_{2}c^{2}}},  \label{rcd} \\
T_{c} &=&\frac{\left( k+m^{2}c_{2}c^{2}\right) ^{3/2}}{3\sqrt{6}\pi q}+\frac{%
m^{2}c_{1}c}{4\pi },  \label{Tcd} \\
P_{c} &=&\frac{\left( k+m^{2}c_{2}c^{2}\right) ^{2}}{96\pi q^{2}}.
\label{Pcd}
\end{eqnarray}%
and we see that for all values of $k$ critical behaviour is possible
provided the constraint
\begin{equation}
k+m^{2}c_{2}c^{2}\geq 0,  \label{Con}
\end{equation}%
is obeyed. Moreover, taking into account Eq. (\ref{EoS}), the pressure is
positive for large volume provided
\begin{equation}
T>\frac{m^{2}c_{1}c}{4\pi },  \label{TT}
\end{equation}%
a relationship that is automatically satisfied provided (\ref{Con}) holds.
Previous investigations of the critical behaviour of black holes in Einstein
gravity have indicated that only spherical topology ($k=1$) admits van der
Waals like behaviour, with the $k=0$ case behaving like an ideal gas. The
graviton mass significantly modifies this behaviour, opening up new
possibilities: topological black holes ($k\neq 1$) can exhibit second order
phase transitions and van der Waals like behaviour (see Fig. \ref{Fig1}).
This admits new prospects for investigating critical behaviour of black
holes in the context of classical gravity, the AdS/CFT correspondence,
holographic interpretation of black holes, and duality with
superconductivity.

\begin{figure}[tbp]
$%
\begin{array}{cc}
\epsfxsize=4cm \epsffile{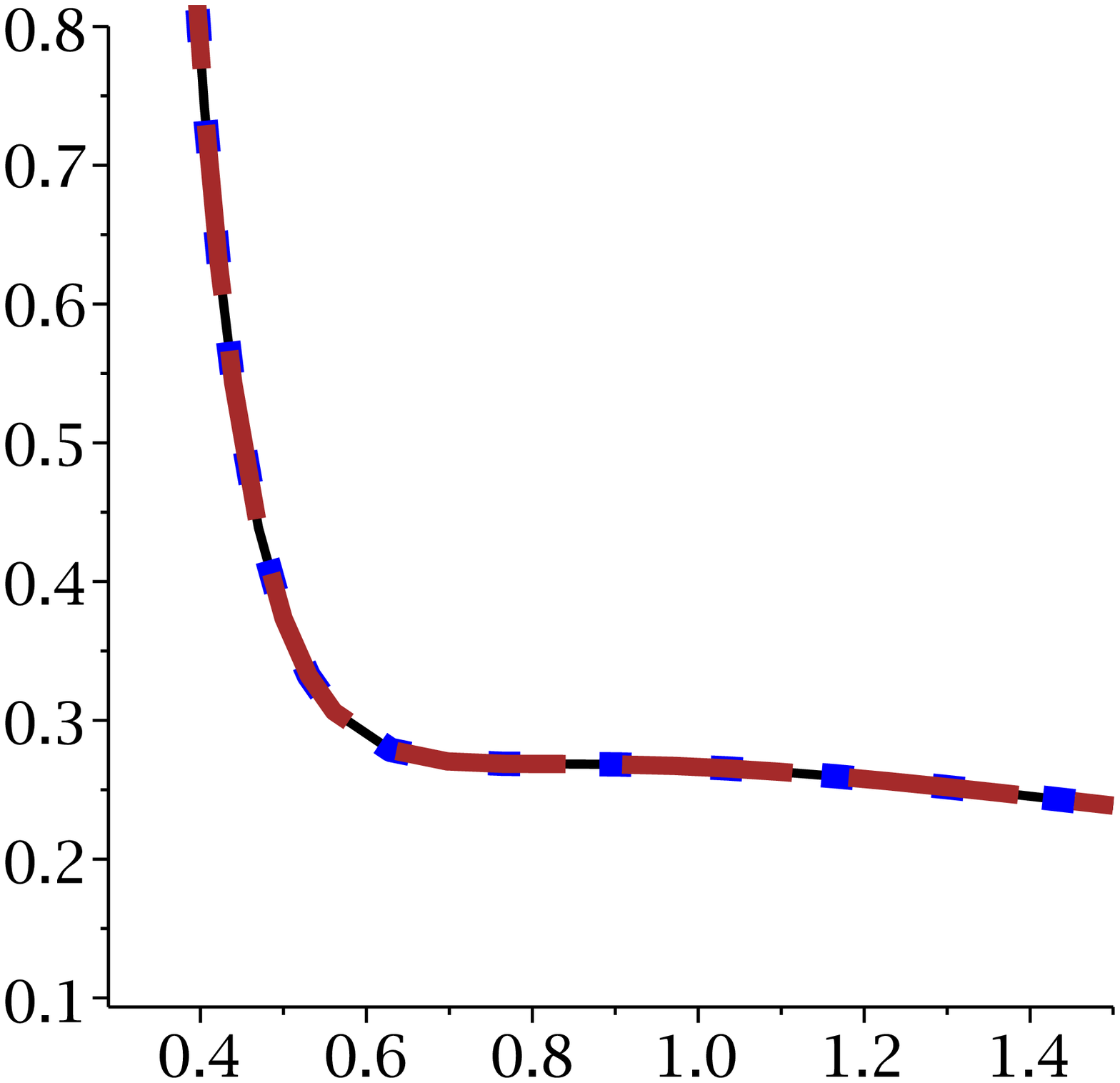} & \epsfxsize=4cm %
\epsffile{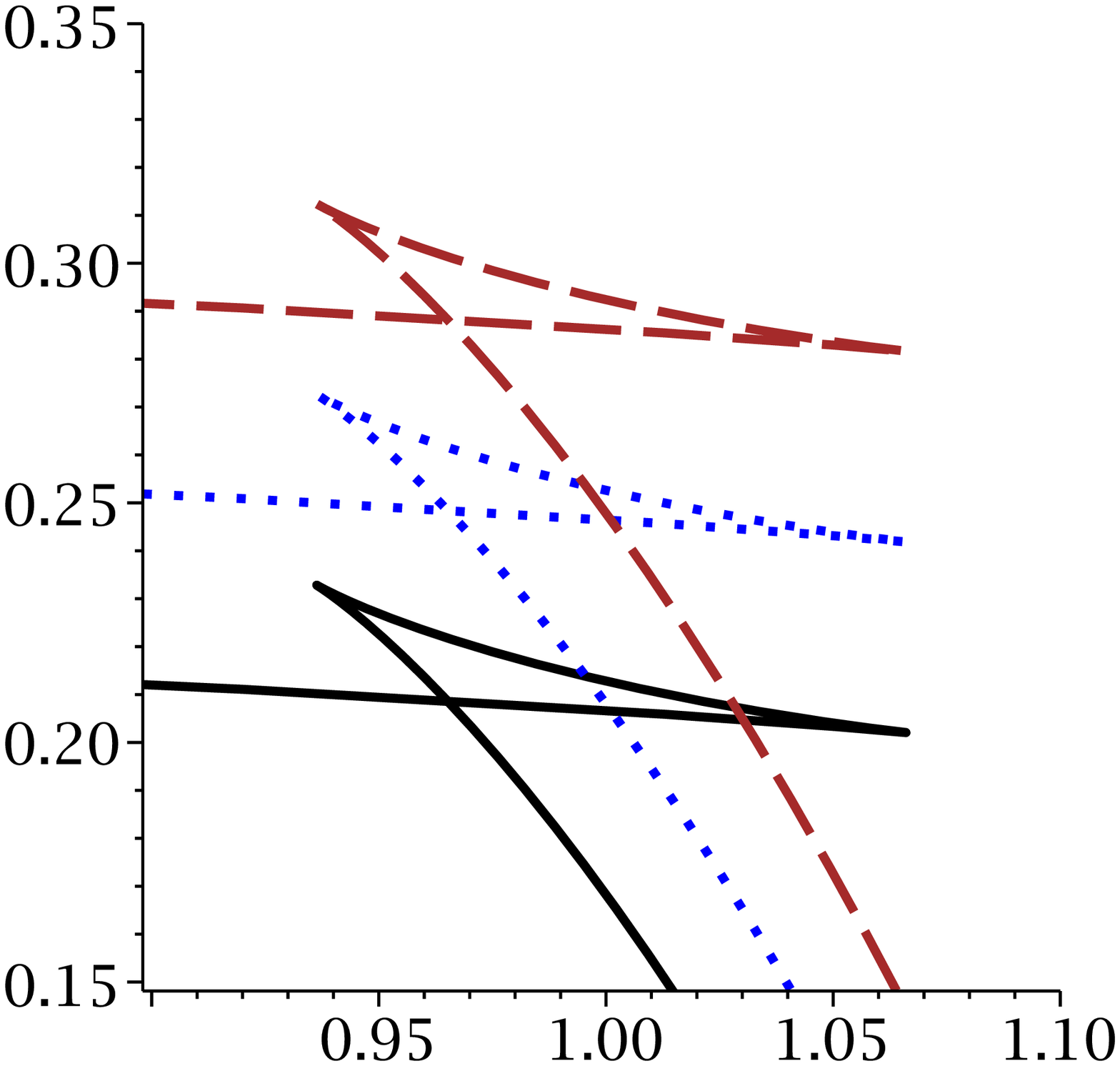}%
\end{array}
$%
\caption{$P-r_{+}$ (left panel) and $G-T$ diagrams for $d=4$,
$q=m=c=c_{1}=1$, $k=-1$ and $c_{2}=10$. Left panel $T=T_{c}$,
right panel $P=0.5P_{c}$: $c_{3}=0$ (continuous line), $c_{3}=1$
(dotted line) and $c_{3}=2$ (dashed line).} \label{Fig2}
\end{figure}

The existence of van der Waals like behaviour for non-spherical
black holes provides another reason for considering modified
versions of Einstein gravity to include massive terms.  In
other words, a non-zero $m$ admits the possibility of critical
behaviour for $k\neq 1$. Furthermore, the massive coefficient
$c_{3}$ in the Gibbs free energy (absent in pressure and
temperature) makes it possible to modify the energy of different
phases, without any modification in critical values and their
corresponding diagrams (see Eq. (\ref{GG}) and Fig. \ref{Fig2}).
For example the formation of a second order phase transition can
take place at the same critical temperature, pressure and horizon
radius but at differing energies of the  various phases.
Such interesting behaviors of Gibbs free energy and critical
values could introduce new phenomenology for the phase structure
of black holes.


\noindent \textit{Closing Remarks-} In this paper we have
demonstrated that topological black holes in dRGT can exhibit van
der Waals behaviour and critical phenomena, in striking contrast
to their counterparts in Einstein gravity. For $k= 0$ it is
sufficient to have any non-zero value of the graviton mass
parameter $m$, whereas for $k=-1$ black holes, this parameter must
be sufficiently large. However too large a value of $m$ will
destabilize the pressure, causing it to become negative for
sufficiently large volume.

Recent  progress in gauge/gravity duality in extended
phase space \cite{Karch}  suggests that massive gravity
will open up avenues of investigation in black hole
thermodynamics. Since such theories admit critical behaviour of
black holes of any horizon curvature, a range of new phenomena in
entanglement entropy \cite{Caceres}, holographic ferromagnetism
\cite{Cai15}, quasinormal modes \cite{Liu}, and
confinement/deconfinment phase transitions for heavy quarks
\cite{Yang} can now be explored.

\medskip \noindent We gratefully thank the anonymous referees for
enlightening comments and suggestions which substantially helped in
improving the quality of the paper. SHH and BE thank Shiraz University
Research Council. The work of RBM was supported by the Natural Science and
Engineering Research Council of Canada. This work has been supported
financially by the Research Institute for Astronomy and Astrophysics of
Maragha, Iran.

\end{document}